\documentclass[printer]{gCOM2e}%
\pdfoutput=1
\newcommand{\pair}{$\ast$-pair}
\newcommand{\shortpaper}[1] {}
\newcommand{\cutforlackofspace}[1] {}
\newcommand{\new}[1] {#1}
\usepackage{graphicx}
\graphicspath{{../}{../Python/}}

\usepackage{url}
\usepackage{subfigure}
\usepackage{doublesp}
\usepackage[nothing]{algorithm}
\usepackage{algorithmic} 
\begin{document}
\doi{10.1080/0003681YYxxxxxxxx}

\markboth{Lemire, Brooks, Yan}{Quasi-Monotonic Segmentation}

\title{An Optimal Linear Time Algorithm for Quasi-Monotonic Segmentation}

\author{Daniel Lemire$^{\ast}$$\dag$\thanks{$^\ast$Corresponding author. Email: lemire@acm.org
\vspace{6pt}}, Martin Brooks, Yuhong Yan${\ddag}$\\\vspace{6pt}  $\dag$University of Quebec at Montreal (UQAM),
 100 Sherbrooke West, Montr\'eal, Qc, H2X 3P2, Canada\\
$\ddag$ National Research Council of Canada,
1200 Montreal Road, Ottawa, ON, Canada, K1A 0R6\\\vspace{6pt}}

\maketitle

\begin{abstract}
Monotonicity is a simple  yet significant qualitative characteristic.  We consider the problem of segmenting a \new{sequence} in up to $K$~segments. We want
segments to be as monotonic as possible and to alternate signs. We propose a quality metric for this problem using the $l_\infty$~norm, and we
present an optimal linear time algorithm based on novel formalism.
Moreover, given a precomputation in time $O(n\log n)$ consisting of a labeling of all extrema, we 
 compute any optimal segmentation in constant time.
We compare experimentally its performance to two piecewise linear 
segmentation heuristics (top-down and bottom-up). We show that our algorithm is faster and more accurate. Applications include pattern recognition and qualitative modeling.
\bigskip

\bigskip

\begin{classcode}H.2.8\end{classcode}
\end{abstract}

\renewcommand{\thefootnote}{}
\footnotetext{This is an expanded version of a conference paper~\cite{YLBICDMO05}.}
\renewcommand{\thefootnote}{\arabic{footnote}}

\section{Introduction}

Monotonicity is one of the most natural and important qualitative properties for sequences of data points. 
It is easy to determine where the values are strictly going up or down, but we only want to
identify significant monotonicity. For example, the drop from 2 to 1.9 in the array $0,1,2,1.9, 3,4$
might not be significant and might even be noise-related. 
 The quasi-monotonic segmentation problem is to determine where the data is approximately increasing or decreasing.
 
 \new{In practical applications, sequences of values can be quite large: it is not
 uncommon to have sensors record data at 10\,kHz or more, thus generating
 terabytes of data and billions of data points. As a dimensionality reduction
 step~\cite{bingham2006}, segmentation divides the data into intervals having homogeneous
 characteristics (flatness, constant slope~\cite{keogh01}, unimodality~\cite{Haiminen04}, monotonicity~\cite{YLBIJCAI05,AAAI05}, step, ramp or impulse~\cite{citeulike:799461}, and so on). The segmentation points can also be used as markers to indicate a qualitative change
 in the data. Other applications include frequent pattern mining~\cite{han98} and  time series classification~\cite{KeoghP98}.
 For qualitative reasoning~\cite{kn:SucECML01}, piecewise monotonic segmentation is especially important as it provides a symbolic model describing system behavior in terms of increasing and decreasing relations between variables.}
 
 \new{There is a trade-off between the number of segments and the approximation error.}
 Some segmentation algorithms~\cite{YLBIJCAI05} give a segmentation having no more than $K$~segments while attempting to minimize the error $\epsilon$; other algorithms~\cite{AAAI05} attempt to minimize the number of segments ($K$) given an upper bound on the error $\epsilon$. We are concerned with the first type of algorithm in this paper.


Using dynamic programming or other approaches, most segmentation problems can be solved in time $O(n^2)$. Other solutions to this problem, using machine learning to classify the pairs of data points~\cite{kn:SucECML01}, are even less favorable since they have higher complexity. However, it is common for sequence of data points to be massive and segmentation algorithms have to have complexity close to $O(n)$ to be competitive. 
 While approximate linear regression segmentation algorithms can be $O(n)$, we show that using a linear regression error to segment according to  monotonicity is not an ideal solution.

We present a metric for the quasi-monotonic segmentation problem
called the Optimal Monotonic Approximation Function Error (OMAFE); this  metric differs
from previously introduced OPMAFE metric~\cite{YLBIJCAI05} since it applies to all segmentations and not just ``extremal'' segmentations. 
We formalize the novel concept of a maximal \pair{} and shows that it can be used to 
define a unique labeling of the extrema leading to an optimal segmentation algorithm.
We also present an optimal linear time algorithm to solve the quasi-monotonic segmentation problem given a segment budget together with an experimental comparison to quantify the benefits of our algorithm. 



\section{Monotonicity Error Metric (OMAFE)}

Finding the best piecewise monotonic approximation can be viewed as
a classical functional approximation problem~\cite{Ubhaya1990}, but we are 
concerned only with discrete  sequences.

Suppose $n$ samples noted $F: D=\{x_1,\ldots, x_n\} \rightarrow \mathbb{R}$ 
with $x_1 < x_2 < \ldots x_n$.
We define, $F_{|[a,b]}$ as the restriction of $F$ over $D\cap [a,b]$. We seek the best monotonic
(increasing or decreasing) function $f:\mathbb{R}\rightarrow \mathbb{R}$ approximating $F$. Let
$\Omega_{\uparrow}$ (resp. $\Omega_{\downarrow}$) be the set of all monotonic increasing (resp. decreasing)
functions. The \textbf{Optimal Monotonic Approximation Function Error (OMAFE)} is $\min_{f \in \Omega}
\max_{x\in D} \vert f(x) - F(x)\vert$ where $\Omega$ is either $\Omega_{\uparrow}$ or $\Omega_{\downarrow}$.

%
%

The segmentation of a set $D$ is a sequence $S = X_1 ,\dots,  X_K$ of intervals in $\mathbb{R}$ with $[\min
D,\max D] = \bigcup_i X_i $ such that $\max X_i =\min X_ {i+1} \in D$ and $X_i \cap X_j= \emptyset$ for
$j \neq i+1, i, i-1$. Alternatively, we can define a segmentation from the set of points $X_i \cap X_
{i+1}=\{y_{i+1}\}$, $y_1= \min X_1$, and $y_{K+1}=\max X_K$. Given $F: \{x_1,\ldots, x_n\} \rightarrow \mathbb{R}$ and a segmentation, the Optimal Monotonic
Approximation Function Error (OMAFE) of the segmentation is  $\max _i \textrm{OMAFE}(F_{|X_i})$ where
the monotonicity type (increasing or decreasing) of the segment $X_i$ is determined by the sign of $F(\max X_i)-F(\min X_i)$. Whenever $F(\max X_i)=F(\min X_i)$, we say the segment has no direction and the best monotonic approximation is just the flat function having value $(\max F_{|X_i} - \min F_{|X_i})/2$. The error is computed over each interval independently; optimal monotonic approximation functions are not required to agree at $\max X_i =\min X_ {i+1}$. Segmentations should alternate between increasing and decreasing, otherwise sequences such as $0,2,1,0,2$ can be segmented as two increasing segments $0,2,1$ and $1,0,2$: we consider it is natural to aggregate segments with the same monotonicity.

We solve for the best monotonic function as follows. 
If we seek the best monotonic increasing function, we first define
$\overline{f}_{\uparrow} (x) = \max \{ F(y): y \leq x \}$ (the maximum of all previous values) and
$\underline{f}_{\uparrow} (x) = \min \{ F(y): y \geq x \}$ (the minimum of all values to come).
 If we seek the best monotonic decreasing function,
we define $\overline{f}_{\downarrow} (x) =  \max \{ F(y): y \geq x \}$ (the maximum of all values to come) and
$\underline{f}_{\downarrow} (x) =\min \{ F(y): y \leq x \}$ (the minimum of all previous values).
These functions, which can
be computed in linear time, are all we need to solve for the best approximation function as shown by the next theorem
which is a well-known result~\cite{Ubhaya1974}.

\begin{theorem}
Given $F: D=\{x_1,\ldots, x_n\} \rightarrow \mathbb{R}$, a best monotonic increasing approximation function to $F$ is $f_{\uparrow}= (\overline{f}_{\uparrow} + \underline{f}_{\uparrow})/2$ and a best monotonic
decreasing approximation function is $f_{\downarrow}=(\overline{f}_{\downarrow} +
\underline{f}_{\downarrow})/2$. The corresponding error (OMAFE) is $\max_{x\in D} (\vert
\overline{f}_{\uparrow}(x) - \underline{f}_{\uparrow}(x) \vert)/2$  or $\max_{x\in D}
(\vert \overline{f}_{\downarrow}(x) - \underline{f}_{\downarrow}(x) \vert)/2$ respectively.
\end{theorem}

The implementation of the algorithm suggested by the theorem is straight-forward.
 Given a segmentation, we can compute the OMAFE in $O(n)$ time
using at most two passes.

\section{A Scale-Based Algorithm for Quasi-Monotonic Segmentation}
\cutforlackofspace{
 The following propositions describe conditions for an optimal segmentation of $F$ if only alternating segmentations are allowed. It sets the stage for a scale-based approach. The next proposition is not used later on but is given as a reference.
 \begin{proposition}\label{stupidresult}
 A segmentation $S=[y_1=x_1,y_2],[y_2,y_3],\ldots,[y_{K},y_{K+1}=x_n]$ of $F: D=\{x_1,\ldots,x_n\} \rightarrow
 \mathbb{R}$ with alternating monotonicity has a minimal number of alternating segments $K$ for a given  $\epsilon$  if for all $i\in \{1,\ldots,K\}$, $[y_i,y_{i+1}]$ has OMAFE less than
 $\epsilon$ (with direction given by $F(y_{i+1})-F(y_i)$) and  $\vert F(y_{i+1})-F(y_i) \vert  \geq 2 \epsilon$.
 \end{proposition}
 \begin{proof}
  Because all segments have OMAFE bounded by $\epsilon$, the segmentation has OMAFE less than $\epsilon$.
   The fact that $K$ is minimal follows
  from  $\vert F(y_{i+1})-F(y_i) \vert  \geq 2 \epsilon$. 
  Suppose you have a better segmentation with fewer segments. Consider $[y_i,y_{i+1}]$: either the interval is partly contained in a same monotonicity segments, or is contained in a different monotonicity segment.  We have that it cannot be contained in a  different monotonicity segment because  $\vert F(y_{i+1})-F(y_i) \vert  \geq 2 \epsilon$ would imply that the OMAFE is at least $\epsilon$ over this interval. One the other hand, if all intervals intersect a new segment with same monotonicity, then the new alternating segmentation has the same number of segments.
 \end{proof}}

 We use the following proposition to prove that the segmentations we generate are optimal (see Theorem~\ref{thm:validlabelling}).
 \begin{proposition}\label{stupidresult2}
 A segmentation $y_1,\ldots,y_{K+1}$ of $F: D=\{x_1,\ldots,x_n\} \rightarrow
 \mathbb{R}$ with alternating monotonicity has a minimal OMAFE $\epsilon$ for a number of alternating segments $K$ if 
\vspace*{-1mm}
 \begin{enumerate}[A.]
\itemsep=0pt\topsep=0pt\partopsep=0pt
\parskip=0pt\parsep=0pt
 \item $F(y_i)=\max F([y_{i-1},y_{i+1}])$ or $F(y_i)=\min F([y_{i-1},y_{i+1}])$ for $i=2,\ldots,K$;
 \item in all intervals $[y_i,y_{i+1}]$  for $i=1,\ldots,K$, there exists $z_1, z_2$ such that $\vert F(z_2)-F(z_1)\vert > 2\epsilon$.
\end{enumerate}
 \end{proposition}
 {\begin{proof}
 Let the original segmentation be the intervals $S_1,\ldots,S_K$ and consider a new segmentation with intervals $T_1, \ldots, T_K$. Assume that the new segmentation has lower error (as given by OMAFE).
 Let $S_i=[y_i, y_{i+1}]$ and $T_i = [y'_i , y'_{i+1}]$.

If any segment $T_m$ contains a segment $S_j$, then the existence of $z_1, z_2$ in $[y_j,y_{j+1}]$ such that $|F(z_2)-F(z_1)| > 2\epsilon$ and $\textrm{OMAFE}(T_m) \le \epsilon$ implies that $T_m$ and $S_j$ have the same monotonicity.

We show that each pair of intervals $S_i$, $T_i$  has nonempty intersection. Suppose not, and let $i$ be the smallest index such that $S_i  \subset  T_{i-1}$.  Since $S_i$ and $T_{i-1}$ have the same monotonicity, for each $j < i$, $S_j$ and $T_j$ have opposite monotonicity. Now consider the $i-1$ intervals $T_1, \ldots, T_{i-1}$  and the $i$ points $y_1, \ldots, y_i$. At least one interval contains two consecutive points; choose the largest  $j < i$ such that $T_j$ contains $y_j, y_{j+1}$.  But then $S_j \subset T_j$, contradicting at least one of the assumptions  $|F(z_2) - F(z_1)| > 2\epsilon$ for $z_1,z_2\in S_i$ and $\textrm{OMAFE}(T_j) \le \epsilon$.

It now follows that each pair of intervals $S_i, T_i$  has the same monotonicity.

Since $\textrm{OMAFE}(T) < \textrm{OMAFE}(S)$, we can choose an index $j$ such that $\textrm{OMAFE}(T_j) < \textrm{OMAFE}(S_j)$. We show that there exists another index $p$ such that $\textrm{OMAFE}(T_p) \ge \textrm{OMAFE}(S_j)$, thus contradicting $\textrm{OMAFE}(T) < \textrm{OMAFE}(S)$.  Suppose $S_j$ is increasing; the proof is similar for the opposite case. Then there exist $x <z \in S_j$ such that $F(x) - F(z) = 2 \times OMAFE(S_j)$. From $\textrm{OMAFE}(T_j) < \textrm{OMAFE}(S_j)$ it follows that at least one of $x$ or $z$ lies in $S_j - T_j$, and hence $F(x) - F(y_j) \ge 2 \times \textrm{OMAFE}(S_j)$ or $F(y_{j+1}) - F(z) \ge 2 \times \textrm{OMAFE}(S_j)$. Thus $\textrm{OMAFE}(T_p) \ge \textrm{OMAFE}(S_j)$ for either $p = j-1$ or $p = j+1$. 
  \end{proof}}

For simplicity, we assume $F$ has no consecutive equal values, i.e.  $F(x_i) \ne F(x_{i+1})$ for $i = 1,\dots, n-1$; our algorithms assume all but one of consecutive equal values values have been removed.
We say  $x_i$ is a maximum if $i \ne 1$ implies $F(x_i) > F(x_{i-1})$ and if $i \ne n$ implies $F(x_i) > F(x_{i+1})$. Minima are defined similarly. 

\begin{figure}
\begin{center}
\includegraphics[width=0.4\columnwidth]{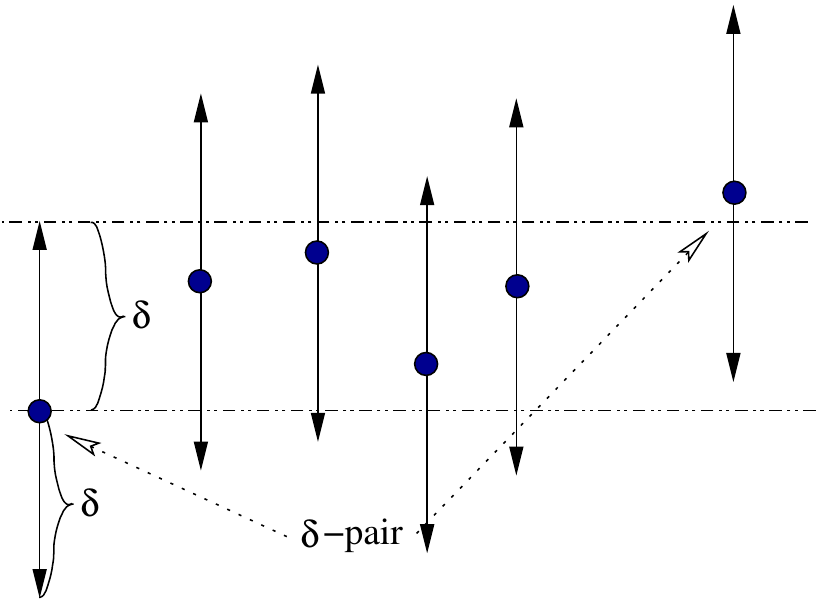}
\end{center}
\caption{\label{deltapair} A $\delta$-pair.  }
\end{figure}

Our mathematical approach is based on the concept of $\delta$-pair{~\cite{brooks94} (see Fig.~\ref{deltapair})}:
\begin{definition}
The tuple $x,y$ ($x<y \in D$) is a $\delta$-pair (or a pair of scale $\delta$) for $F$ if $|F(y)-F(x)| \geq \delta$ and for
all $z \in D$, $x<z<y$ implies $|F(z)-F(x)|< \delta$ and $|F(y)-F(z)|< \delta$. A $\delta$-pair's \textit{direction} is \textit{increasing} or \textit{decreasing}
according to whether $F(y)>F(x)$ or $F(y)<F(x)$.
\end{definition}
  $\delta$-Pairs having opposite directions cannot overlap but they may share an end~point.
$\delta$-Pairs of the same direction may overlap, but may not be nested. We use the term ``\pair{}'' to
indicate a $\delta$-pair having an unspecified $\delta$. We say that a \pair{} is significant at scale $\delta$ if it is of scale $\delta'$ for $\delta' \geq \delta$.
 From a topological viewpoint, a \pair{} is the pairing of critical
points used to determine each extremum's persistence \cite{Edelsbrunner2002}.

 We define $\delta$-monotonicity as follows:
 \begin{definition}
 Let $X$ be an interval, $F$ is $\delta$-\textit{monotonic} on $X$ if all
 $\delta$-pairs in $X$ have the same direction; $F$ is \textit{strictly} $\delta$-monotonic when there exists at least one such $\delta$-pair. In this case:
\vspace*{-1mm}
 \begin{itemize}
\itemsep=0pt\topsep=0pt\partopsep=0pt
\parskip=0pt\parsep=0pt
 \item $F$ is $\delta$-\textit{increasing} on $X$ if $X$ contains an increasing
 $\delta$-pair.
 \item $F$ is $\delta$-\textit{decreasing} on $X$ if $X$ contains a decreasing
 $\delta$-pair.
 \end{itemize}
  \end{definition}

A $\delta$-monotonic interval $X$ satisfies $\textrm{OMAFE}(X)<\delta/2$. 
We say that a \pair{} $x,y$ is \textbf{maximal} if whenever $z_1,z_2$ is a \pair{} of a larger scale in the same direction containing $x,y$, then there exists a \pair{} $w_1,w_2$ of an opposite direction contained in $z_1,z_2$ and containing $x,y$.  For example, the sequence $1,3,2,4$ has 2 maximal \pair{}s: $1,4$ and $3,2$. 
  Maximal \pair{}s of opposite direction may share a common point, whereas maximal \pair{}s of the same direction may not. Maximal \pair{}s cannot overlap, meaning that it cannot be the case that exactly one end point of a maximal \pair{} lies strictly between the end points of another maximal  \pair{}; either neither point lies strictly between or both do. In the case that both do, we say that the one maximal \pair{} properly contains the other. All \pair{}s must be contained in a maximal \pair{}.

 \begin{lemma}
 The smallest maximal \pair{} containing a \pair{} must be of the same direction.
 \end{lemma}
 
\begin{proof} Suppose a \pair{} is immediately contained in a maximal \pair{} $W$. Suppose $W$ is not
in the same direction, then within $W$, seek the largest \pair{} in the
same direction as $P$ and containing $P$, then it must be a maximal \pair{} in $D$ since maximal
\pair{}s of different directions cannot overlap.\end{proof}

The first and second point of a maximal \pair{} are extrema and the reverse is true as well as shown by the next lemma. 

\begin{lemma}\label{prop:either}
Every extremum is either the first or second point of a maximal \pair{}.
\end{lemma}
\begin{proof}
The case $x=x_1$ or $x=x_n$ follows by inspection. Otherwise, $x$ is the end point of a left and a right \pair{}. Each \pair{} must immediately belong to a maximal \pair{} of same direction: a \pair{} $P$ is contained in a maximal \pair{} $M$ of same direction and there is no maximal \pair{} $M'$ of opposite direction such that $P\subset M'\subset M$. Let $M^l$ and $M^r$ be the maximal \pair{}s immediately containing the left and right \pair{} of $x$. Suppose neither $M^l$ and $M^r$ have $x$ as a end point. Suppose $M^l \subset M^r$, then the right \pair{} is not immediately contained in $M^r$, a contradiction.  The result follows by symmetry.
\end{proof}


Our approach is to label each extremum in $F$ with a scale parameter $\delta$ saying that this extremum is ``significant'' at scale $\delta$ and below.
%
Our intuition is that by picking extrema at scale $\delta$, we should have a segmentation having error less than $\delta/2$.

\begin{definition}\label{def:optscalelabel}The scale labeling of an extremum $x$ is the maximum of the scales of the maximal \pair{}s for which it is an end point.
\end{definition}

For example, given the sequence $1,3,2,4$ with  2 maximal \pair{}s ($1,4$ and $3,2$), we would give the following labels in order $3,1,1,3$.

\begin{definition}Given $\delta > 0$, a {\it maximal alternating sequence of $\delta$-extrema} 
$Y = y_1 \dots y_{K+1}$ 
is a sequence of extrema each having scale label at least $\delta$, having alternating types (maximum/minimum), and such that there exists no
sequence properly containing $Y$ having these same properties. From $Y$ we define a
{\it maximal alternating $\delta$-segmentation} of $D$ by segmenting at the points $x_1, y_2 \dots y_K, x_n$. 
\end{definition}

\begin{theorem}\label{thm:validlabelling}
Given $\delta > 0$, let $P = S_1 \dots S_K$ be a maximal 
alternating $\delta$-segmentation derived from maximal alternating sequence $y_1 \dots y_{K+1}$ of $\delta$-extrema.
Then any alternating segmentation $Q$ having OMAFE($Q$) $<$ OMAFE($P$) has at least $K+1$ segments.
\end{theorem}

\begin{proof} We show that conditions A and B of Proposition~\ref{stupidresult2} are satisfied with $\epsilon =$ OMAFE($P$).

First we show that each segment $S_i$ is $\delta$-monotone; from this we conclude that $\textrm{OMAFE}(P) < \delta/2$.
Intervals $[x_1,  y_1]$ and $[y_K, x_n]$ contain no maximal \pair{}s of scale $\delta$ or larger, and
therefore contain no \pair{}s of scale $\delta$ or larger. Similarly, no $[y_i, y_{i+1}]$ contains an opposite-direction significant
\pair{}.

Condition A: Follows from $\delta$-monotonicity of each $S_i$ and maximal \pair{}s not overlapping.

Condition B: We show that $\vert F(y_{i+1})-F(y_i) \vert  \geq \delta > 2 \times \textrm{OMAFE}(P)$. If $i=1$, then $y_i$ must begin an maximal \pair{}, and the maximal \pair{} must end with $y_{i+1}$ since maximal \pair{}s cannot overlap. The case $i+1=k$ is similar. Otherwise, since maximal \pair{}s cannot overlap, each $y_i, y_{i+1}$ is either  a maximal \pair{} of scale $\delta$ or larger or there exist indices $j$ and $k$, $j < i$  and $k > i+1$ such that $y_j , y_i$ is a maximal \pair{} of scale at least $\delta$, and $y_{i+1}, y_k$ is a maximal \pair{} of scale at least $\delta$. These two maximal \pair{}s have the same direction, and that this is opposite to the direction of $[y_i  y_{i+1}]$. 
Now suppose  $\vert F(y_i) - F(y_{i+1})| < \delta$. Then $y_j, y_k$ is a \pair{} properly containing $y_j$, $y_i$ and $y_{i+1}$, $y_k$. 
But neither $y_j , y_i$ nor $y_{i+1}, y_k$ can be properly contained in a \pair{} of opposite direction lying within
$y_j, y_k$, thus contradicting their maximality and proving the claim. 
\end{proof}

Sequences of extrema labeled at least $\delta$ are generally not maximal alternating. For example the sequence 
$0,10,9,10,0$ is scale labeled $10,10,1,10,10$.
However, a simple relabeling of certain extrema can make them maximal alternating. Consider two same-sense extrema $z_1 < z_2$
such that lying between them there exists no extremum having scale at least as large as the minimum of the two extrema's scales. We must have $F(z_1) = F(z_2)$, since otherwise the point upon which $F$
has the lesser value could not be the endpoint of a maximal \pair{}.
This is the only situation which causes choice when constructing a maximal alternating sequence of $\delta$-extrema.
To eliminate this choice, replace the scale label on $z_1$ with the largest scale of the opposite-sense extrema lying between them.
{In the next section, Algorithm~\ref{algo:computedelta} incorporates this re-labeling making Algorithm~\ref{algo:sortdelta} simple and efficient. }

\cutforlackofspace{

In light of Proposition~\ref{stupidresult2}, the next proposition says that the significant extrema gives an optimal segmentation except when there are successive maxima or minima as in this sequence  $0,10,9,10,-1$ at a scale strictly greater than 1.

\begin{proposition}\label{prop:validlabelling}
If for a given $\delta$, all significant
extrema of $F: D=\{x_1,\ldots,x_n\} \rightarrow \mathbb{R}$ at scale $\delta$, $\{y_1,\ldots, y_K\}\subset D$, are
such that there are no successive maxima or successive minima then
\vspace*{-1mm}
\begin{itemize}
\itemsep=0pt\topsep=0pt\partopsep=0pt
\parskip=0pt\parsep=0pt
\item $F$ is $\delta$-monotonic on $[y_i, y_{i+1}]$ for all $i$;
\item $\vert F(y_{i+1})-F(y_i) \vert  \geq \delta$ for all $i$;
\item $F(y_i)= \max F([y_{i-1},y_{i+1}])$ or $F(y_i)= \min F([y_{i-1},y_{i+1}])$ for all $i$; 
\item there are no \pair{} of scale $\delta$ or more in $[x_1,y_1]$ or in $[y_{K-1}, x_n]$.
\end{itemize}
\end{proposition}
\begin{proof}
Suppose that $[y_i,y_{i+1}]$ is not $\delta$-monotonic. For simplicity, suppose $y_i$ is a maximum and $y_{i+1}$ is a minimum. We have that $[y_i,y_{i+1}]$ must
contain an increasing $\delta'$-pair for $\delta'\geq\delta$ (otherwise, it is $\delta$-monotonic) which implies we omitted a significant minimum followed by a maximum inside $[y_i,y_{i+1}]$, a contradiction.

Next we show that $\vert F(y_{i+1})-F(y_i) \vert  \geq \delta$. If $i=1$, then $y_i$ must begin an maximal \pair{}, and the maximal \pair{} must end with $y_{i+1}$ since maximal \pair{}s cannot overlap. The case $i+1=k$ is similar. Otherwise, since maximal \pair{}s cannot overlap, each $y_i, y_{i+1}$ is either  a maximal \pair{} of scale $\delta$ or larger or there exist indices $j$ and $k$, $j < i$  and $k > i+1$ such that $y_j , y_i$ is a maximal \pair{} of scale at least $\delta$, and $y_{i+1}, y_k$ is a maximal \pair{} of scale at least $\delta$. 
These two maximal \pair{}s have the same direction, and that this is opposite to the direction of $[y_i  y_{i+1}]$. 
Now suppose  $\vert F(y_i) - F(y_{i+1})| < \delta$. Then $y_j, y_k$ is a \pair{} properly containing $y_j$, $y_i$ and $y_{i+1}$, $y_k$. 
But neither $y_j , y_i$ nor $y_{i+1}, y_k$ can be properly contained in a \pair{} of opposite direction lying within
$y_j, y_k$, thus contradicting their maximality and proving the claim.

Suppose there is a \pair{} of scale at least $\delta$ in $[x_1,y_1]$, then we have that both end~points must be significant extrema contradicting the fact that $y_1$ is the first significant extrema. Other results follow by inspection.
\end{proof}

 The next proposition shows that when we have successive significant maxima or successive significant minima, then they have the same value.
 
\begin{proposition}\label{prop:repeatedextrema}
For a given $\delta$, if $y,z\in D$ are
successive significant maxima or successive significant minima, then $F(y)=F(z)$.
\end{proposition}
\begin{proof}
If there is no significant extrema between $y,z\in D$, then there is no significant maximal \pair{} in $[y,z]$. Because $y$ and $z$ are significant. there exists $y'$ and $z'$ such that $y',y$ and $z,z'$ are significant maximal \pair{}s of opposite direction. If $F(y) \neq F(z)$, we can extend either $y',y$ or $z,z'$, a contradiction.
\end{proof}

For an optimal segmentation, it suffices to prune successive maxima or minima and we call the result of such a pruning a $\delta$-segmentation.

}


\cutforlackofspace{
In light of Proposition~\ref{prop:unique}, we note the valid labeling of an extremum $x$ of $F$ to be $\delta_F(x)$. We will define an \textit{optimal labeling} to be a valid labeling with repeated extrema as a special case.
 
 \begin{definition}Given a function $F$ choose any sequence of functions $F^{(i)}$ with no two extrema have the same value such that $F^{(i)} \rightarrow F$ pointwise over D. If $\Delta_F(x)=\lim_{i\rightarrow \infty} \delta_{F^{(i)}}(x)$ exists for all extrema $x$ of $F$, then $\Delta_F(x)$ is an optimal labeling.\end{definition}

An optimal labeling always exists.
 }
 
\cutforlackofspace{ 
To ensure that the significant extrema form a $\delta$-segmentation, we simply relabel some of the extrema having the same value. Given any two maxima $x_1$ and $x_2$ (resp. minima) at scale $\delta_1$ and $\delta_2$ such that the right maximal \pair{} of $x_1$ is at scale $\delta'<\min (\delta_1, \delta_2)$, then we label $x_1$ with $\delta'$.
  The following result follows from Propositions \ref{stupidresult2} and \ref{prop:validlabelling}.

\begin{theorem}\label{oldthm:validlabelling}
For a given $\delta$, any $\delta$-segmentation is an optimal segmentation having OMAFE less than $\delta/2$: no segmentation having the same number of alternating segments can have a smaller OMAFE.
\end{theorem}

Hence, given a budget of $K$ segments, and pick $\delta$ as small as possible so that you have no more than $K+1$ significant extrema, and you have an optimal segmentation. Also, should the original data be already piecewise monotonic, picking the right number of monotonic segments $K$ leads to a perfect segmentation. 
}

\subsection{Computing a Scale Labeling Efficiently}

Algorithm~\ref{algo:computedelta} (next page) produces a scale labeling in linear time. Extrema from the original data are visited in order, and they alternate (maxima/minima) since we only pick one of the values when there are  repeated values (such as $1,1,1$).  

The algorithm has a main loop (lines 5 to 12) where it labels extrema as it identifies extremal \pair{}s, and stack the extrema it cannot immediately label.  At all times, the stack (line 3) contains minima and maxima in \textbf{strictly} increasing and decreasing order respectively. Also at all times, the last two extrema at the bottom of the stack are the absolute maximum and absolute minimum (found so far).  Observe that we can only label an extrema as we find new extremal \pair{}s (lines 7, 10, and 14).
\vspace*{-1mm}
\begin{itemize}
\itemsep=0pt\topsep=0pt\partopsep=0pt
\parskip=0pt\parsep=0pt
\item If the stack is empty or contains only one extremum, we simply add the new extremum (line 12).

\item If there are only 2 extrema $z_1,z_2$ in the stack and we found either a new absolute maximum or new absolute minimum ($z_3$), we can pop and label the oldest one ($z_1$) (lines 9, 10, and 11) because the old pair ($z_1,z_2$) forms a maximal \pair{} and thus must be bounded by extrema having at least the same scale while the oldest value ($z_1$) does not belong to a larger maximal \pair{}. 
Otherwise, if there are only 2 extrema $z_1,z_2$ in the stack and the new extrema $z_3$ satisfies $z_3 \in (\min(z_1,z_2), \max(z_1,z_2))$, then we add it to the stack since no labeling is possible yet.

\item While the stack contains more than 2 extrema (lines 6, 7 and 8), we consider the last three points on the stack ($s_3,s_2,s_1$) where $s_1$ is the last point added. Let $z$ be the value of the new extrema. If $z \in (\min(s_1,s_2), \max(s_1,s_2))$, then it is simply added to the stack since we cannot yet label any of these points; we exit the while loop. Otherwise, we have a new maximum (resp. minimum) exceeding (resp. lower) or matching the previous one on stack, and hence $s_1,s_2$ is a maximal \pair{}. If $z\neq s_2$, then $s_3, z$ is a maximal \pair{} and thus, $s_2$ cannot be the end of a maximal \pair{} and $s_1$ cannot be the beginning of one, hence both $s_2$ and $s_1$ are labeled. If $z=s_2$ then we have successive maxima or minima and the same labeling as $z\neq s_2$ applies.
\end{itemize}
During the ``unstacking'' (lines 13 and following), we visit a sequence of minima and maxima 
forming increasingly larger maximal \pair{}s.

The algorithm runs in time $O(n) $ (independent of $K$). Indeed, for any index of an extremum, the condition at line~\ref{whileline} will evaluate once to false; moreover the condition at line~\ref{whileline} cannot evaluate to true more than $O(n)$ times.

\begin{algorithm}

 \begin{algorithmic}[1]
 \STATE \textbf{INPUT:} an array $d$ containing the $y$ values indexed from $0$ to $n-1$, repeated consecutive values have been removed
 \STATE \textbf{OUTPUT:} a scale labeling for all extrema
 \STATE $S\leftarrow$ empty stack, First($S$) is the value on top, Second($S$) is the second value
 \STATE \textbf{define} $\delta(d,S)=\vert d_{\textrm{First}(S)}-d_{\textrm{Second}(S)}\vert$
 \FOR {$e$ index of an extremum in $d$, $e$'s are visited in increasing order}
 \WHILE{length($S$) $>2$ and ($e$ is a minimum such that $d_e\leq \textrm{Second}(S)$ or $e$ is a maximum such that $d_e\geq \textrm{Second}(S)$)}\label{whileline}
 \STATE label First($S$) and Second($S$) with $\delta(d,S)$
 \STATE pop stack $S$ twice
 \ENDWHILE
 \IF{length($S$) is 2 and ($e$ is a minimum such that $d_e\leq \textrm{Second}(S)$ or $e$ is a maximum such that $d_e\geq \textrm{Second}(S)$)}
  \STATE label Second($S$) with $\delta(d,S)$
  \STATE remove Second($S$) from stack $S$
 \ENDIF
 \STATE stack $e$ to $S$
 \ENDFOR
 \WHILE{length of $S$ $>2$}
 \STATE  label First($S$) with $\delta(d,S)$
 \STATE pop stack $S$
 \ENDWHILE
 \STATE label First($S$) and Second($S$) with $\delta(d,S)$
 \end{algorithmic}
\caption[Algorithm computing a scale labeling.]{\label{algo:computedelta}Algorithm to compute the scale labeling in $O(n)$ time. 
}
\end{algorithm}

Once the labeling is complete, we find $K+2$ extrema having largest scale in time $O(n K)$ using $O(K)$ memory, then we remove all extrema having the same scale as the smallest scale in these $K+2$ extrema (removing at least one), we replace the first and the last extrema by $0$ and $n-1$ respectively (see Algorithm~\ref{algo:sortdelta}). The result is an optimal segmentation having at most $K$ segments.

 \begin{algorithm}
  \begin{algorithmic}
  \STATE \textbf{INPUT:} an array $d$ containing the $y$ values indexed from $1$ to $n$
  \STATE \textbf{INPUT:} $K$ a bound on the number of segments desired
  \STATE \textbf{OUTPUT:} unsorted segmentation points (a $\delta$-segmentation)
  \STATE  $L \leftarrow $ empty array (capacity $K+3$)
  \FOR {$e$ is index of an extremum in $d$ having scale $\delta$, $e$ are visited in increasing order}
    \STATE insert $(e,\delta)$ in $L$ so that $L$ is sorted by scale in decreasing order (sort on $\delta$) using binary search
    \IF{length of $L$ is $K+3$}
      \STATE pop last(L)
    \ENDIF
  \ENDFOR
  \STATE remove all elements of $L$ having the scale of last(L)
  \STATE \textbf{RETURN:} the indexes in $L$ replacing first one by $1$ and last one by $n$
  \end{algorithmic}
 \caption[Algorithm returning a segmentation using at most $K$ segments.]{\label{algo:sortdelta}Given the scale labeling, this algorithm will return a segmentation using at most $K$ segments. It is assumed that there are at least $K+1$ extrema to begin with.}
 \end{algorithm}

Alternatively, if we plan to resegment the time series
several times with different values of $K$, 
we can sort all extrema by their label in time $O(n\log n)$, and 
compute in time $O(n)$ an auxiliary structure on the sorted set so that when selecting the $i\textrm{th}$~item
in the sorted list ($d_i$), we obtain the index $j$ of the earliest occurrence of this scale in the
list ($\textrm{scale}(d_j)=\textrm{scale}{d_i}$ and $\textrm{scale}(d_j)<\textrm{scale}(d_{j-1})$ if $j>0$) in constant time.
Hence, we can segment any time series optimally in constant time
given this precomputation in time  $O(n\log n)$.

\begin{lemma}\label{oldthm:validlabelling}
Given a precomputation in time $O(n \log n)$ using $O(n)$~storage,
for any desired upper bound on the number of segments $K$, we can compute the segmentation points of an optimal OMAFE, and the corresponding
OMAFE value, in constant time.
\end{lemma}

Hence, we can compute an OMAFE versus $K$ plot in $O(n \log n)$ time.

\begin{algorithm}[b]
 \begin{algorithmic}
\STATE \textbf{INPUT:} Time Series $(x_i,y_i)$ of length $n$
\STATE \textbf{INPUT:} Desired number of segments $K$
\STATE \textbf{INPUT:} Function $E(p,q)$ computing linear fit error in range $[x_p,x_q]$
 \STATE $S\leftarrow (1,n, E(0,n))$
 \WHILE{ $\vert S \vert < K$ }
\STATE find tuple $(i,j,\epsilon)$ in $S$ with maximum last entry 
\STATE find minimum of $E(i,l)+E(l+1,j)$ for $l=i,\ldots,j-1$
\STATE remove tuple $(i,j,\epsilon)$ from $S$
\STATE insert tuples $(i,l,E(i,l))$ and $(l,j,E(l+1,j))$ in $S$
 \ENDWHILE
\STATE $S$ contains the segmentation
 \end{algorithmic}
\caption{\label{algo:topdown}Piecewise Linear Top-Down Segmentation Heuristic.}
\end{algorithm}

\begin{algorithm}
 \begin{algorithmic}
\STATE \textbf{INPUT:} Time Series $(x_i,y_i)$ of length $n$
\STATE \textbf{INPUT:} Desired number of segments $K$
\STATE \textbf{INPUT:} Function $E(p,q)$ computing linear fit error in range $[x_p,x_q]$
\STATE $S\leftarrow [0,0],[1,1],[2,2],\ldots,[n,n] $ 
 \WHILE{ $\vert S \vert > K$ }
\STATE find consecutive intervals in $S$, $[p_1,p_2]$ and $[p_2+1,p_3]$, having minimal value $E(p_1,p3)-E(p_1,p_2)-E(p_2+1,p_3)$
\STATE merge the two consecutive intervals
 \ENDWHILE
\STATE $S$ contains the segmentation
 \end{algorithmic}
\caption{\label{algo:bottomup}Piecewise Linear Bottom-Up Segmentation Heuristic.}

\end{algorithm}

\section{Experimental Results and Comparison to Piecewise Linear Segmentation Heuristics}\label{experimental}

\new{We compare our optimal $O(n K)$ algorithm with  our implementations of two piecewise linear
segmentation heuristics~\cite{keogh01}:
  top-down, which runs in $O(n K)$ time (see Algorithm~\ref{algo:topdown}), and
  bottom-up which runs in $O(n(n-K))$~time
 (see Algorithm~\ref{algo:bottomup})}. The top-down heuristic
successively segments the data starting with only one segment, each time picking the segment with the worse
linear regression error and finding the best segmentation point; the linear regression is not continuous from one segment to the other. The regression error can be computed in
constant time if one has precomputed the range moments~\cite{LemireCASCON2002,LemireSDM2007}. 
\new{The bottom-up heuristic
starts with intervals containing only one data point and successively merge them,
each time choosing the least expensive merge. By maintaining the segments in a 
doubly-linked list coupled with a heap or tree, it is possible to obtain
a bottom-up heuristic with $O((n-K) \log n)$~complexity, but it then uses much more 
memory and it is more difficult to implement.}

Once the piecewise linear
segmentation is completed, we run through the
segments and aggregate consecutive segments having the same sign where the sign of a segment $[y_k,y_{k+1}]$ is defined by  $F(y_{k+1})-F(y_{k})$, setting 0 to be a positive sign (increasing monotonicity).

\begin{figure}
    \center{
    \includegraphics[width=0.4\columnwidth]{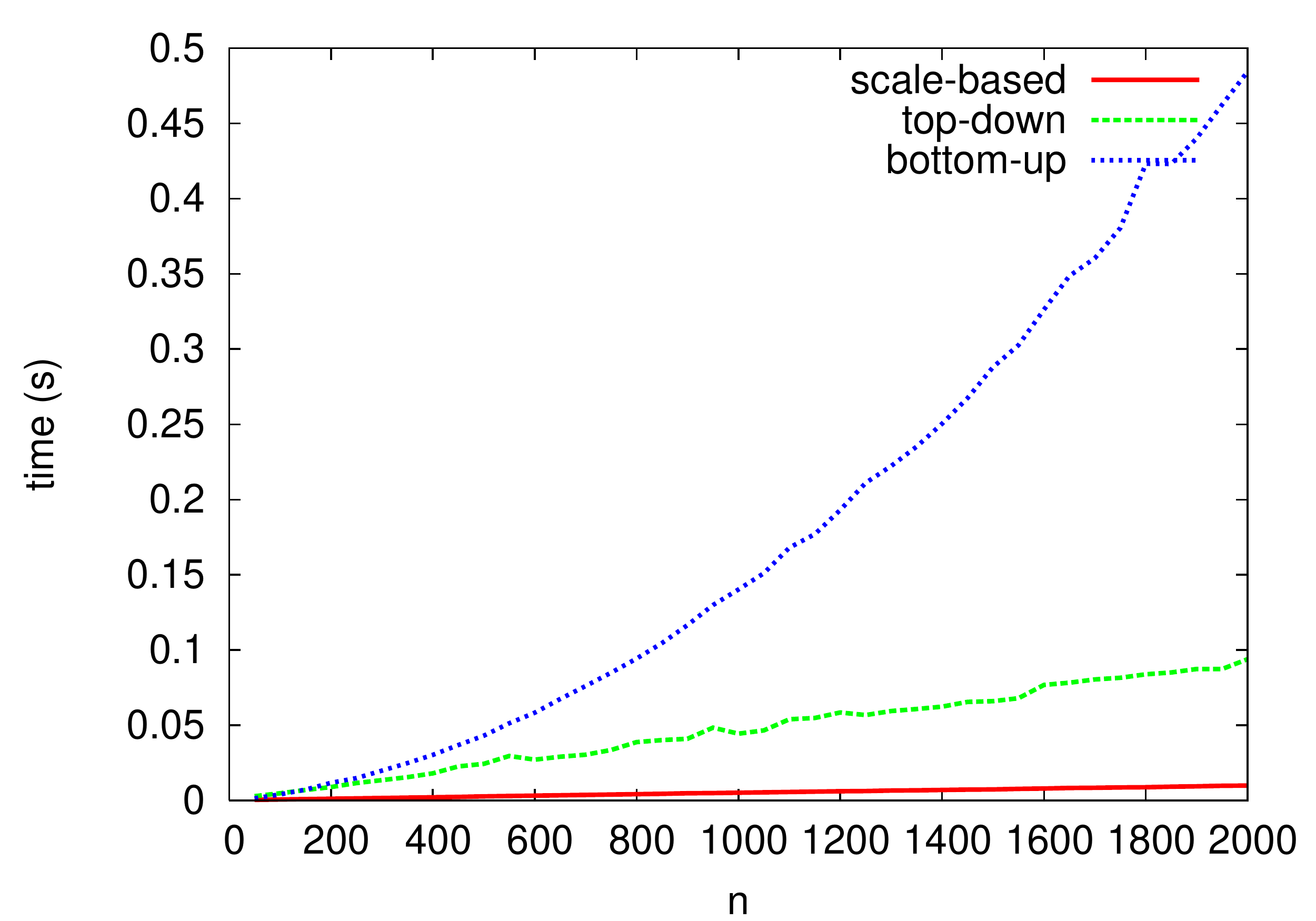}
    }

\caption{\label{relativetimings}Time to segment a time series of length $n$ in $K=20$ segments.}
\end{figure}

\new{We implemented all algorithms in
Python (version 2.5) and ran the experiments on a 2.16\,GHz Intel Core 2 Duo processor with sufficient RAM (1\,GB).  
 Fig.~\ref{relativetimings}
presents the relative speed of the various segmentation algorithms on time series
of various lengths for a fixed number of segments (using randomly generated data). The timings reported include
all pre-processing.}

\subsection{Electrocardiograms (ECG)}

\begin{figure}
    \center{
    \includegraphics[width=0.4\columnwidth]{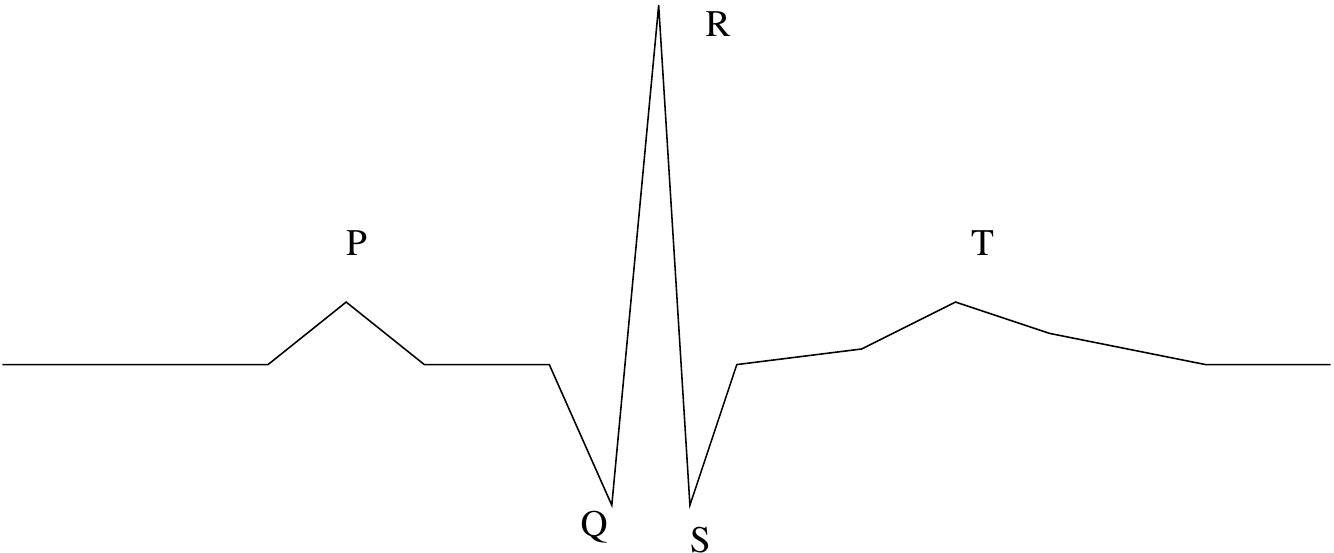}
    }

\caption{\label{pqrst}Schema of an ECG pulse with commonly identified reference points (PQRST).}
\end{figure}

\begin{figure}

\subfigure[\label{basicecg}Time Series]{
\centering\includegraphics[width=0.33\columnwidth]{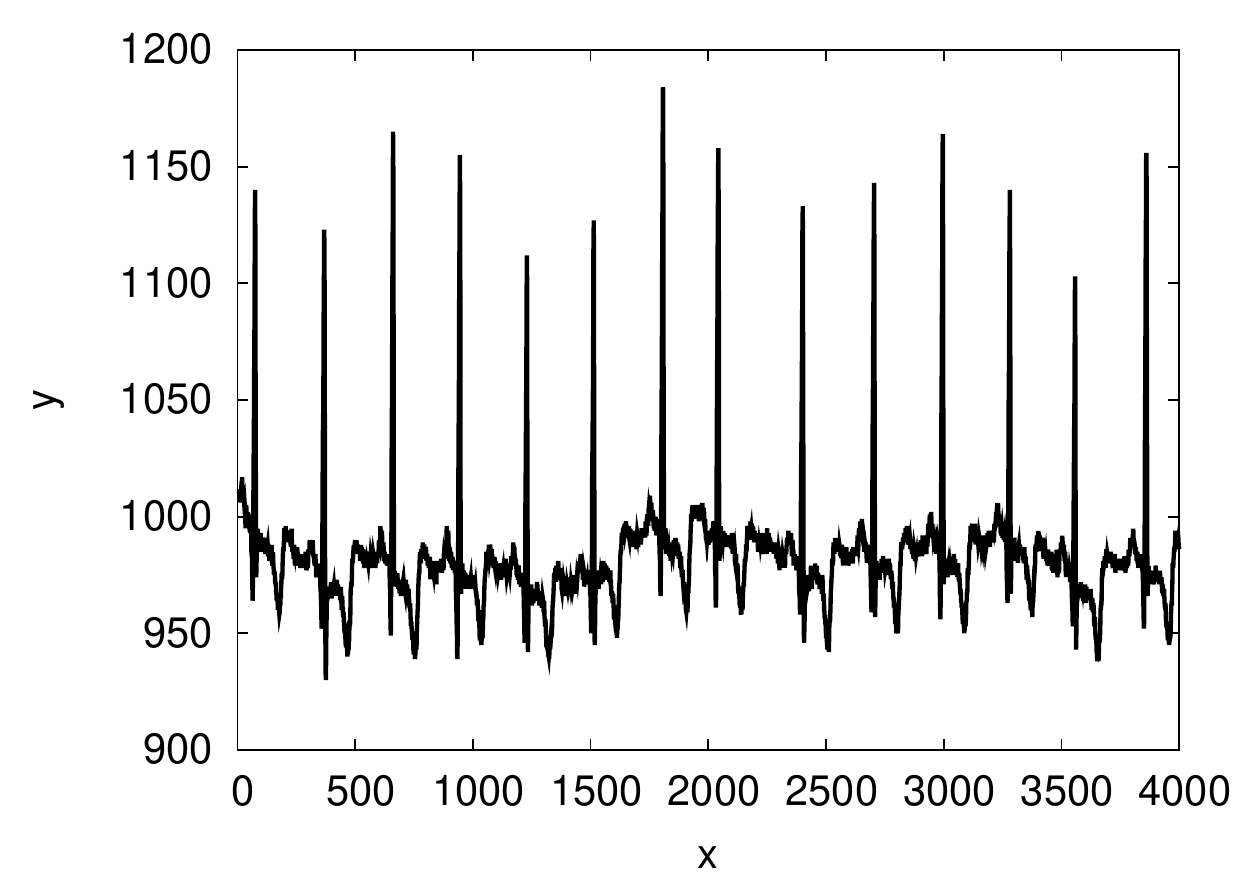}
}
\subfigure[\label{timevsk}Time vs. number of segments $K$]{
\centering\includegraphics[width=0.33\columnwidth]{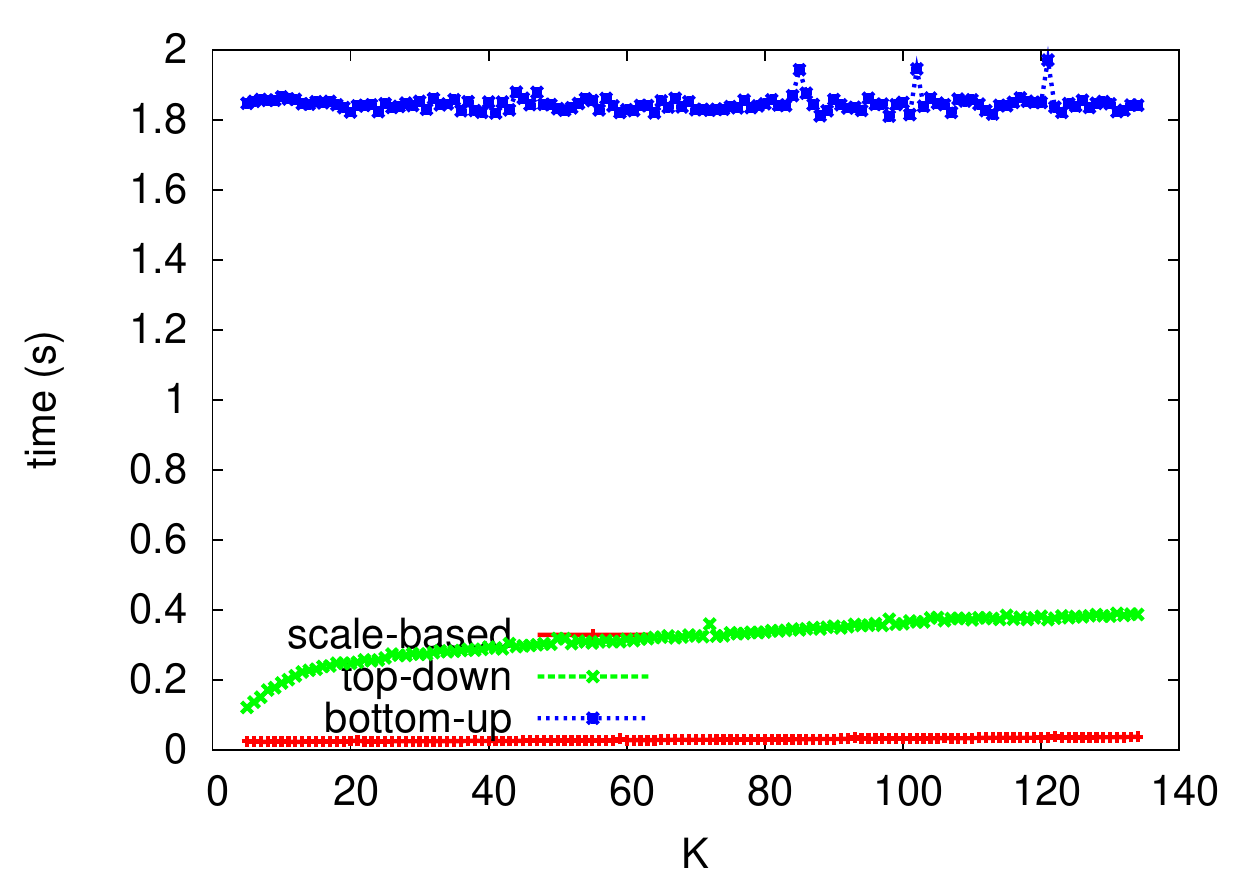}
}
\subfigure[\label{omafevsk}OMAFE vs. number of segments $K$]{
\centering\includegraphics[width=0.33\columnwidth]{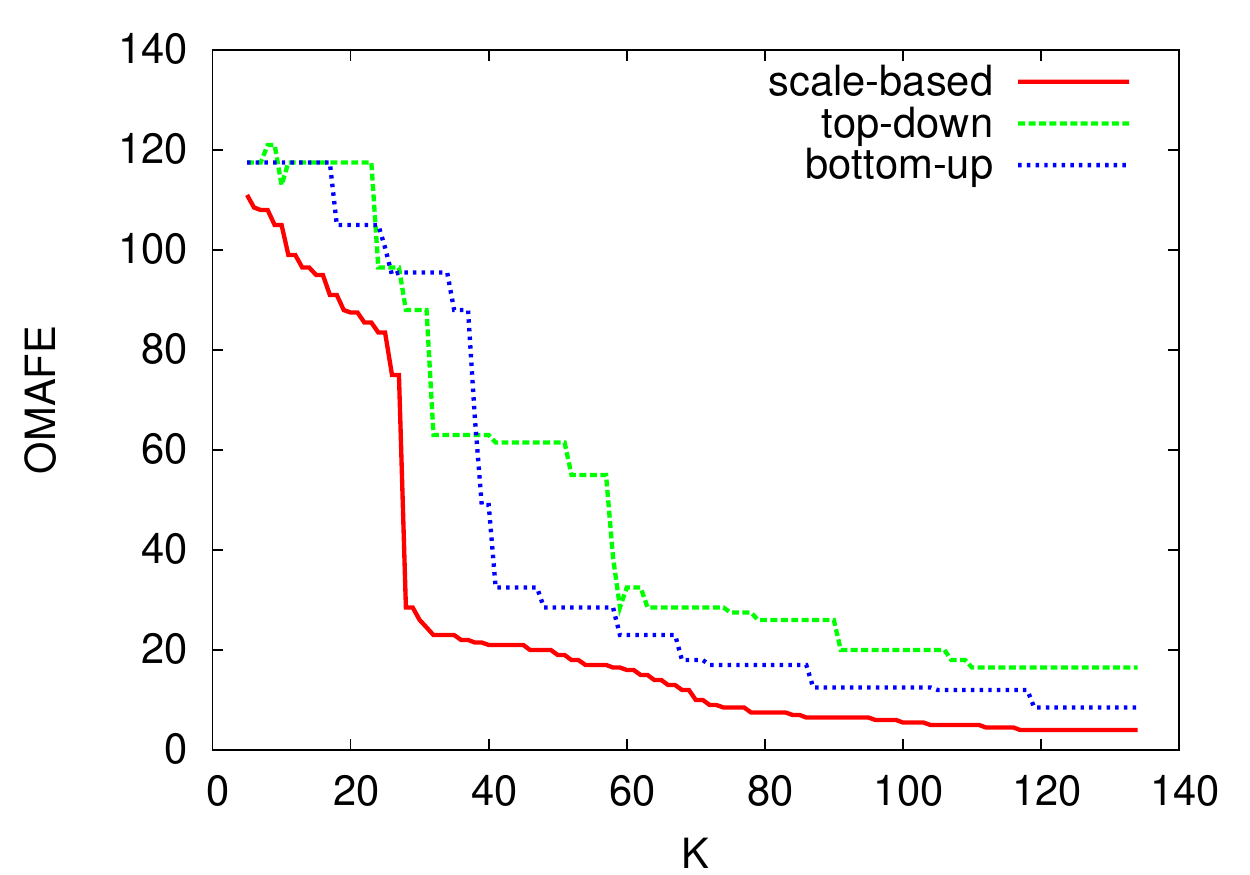}
}
\caption{Results of experiments over ECG data.}
\end{figure}

 ECGs have a well known monotonicity structure with 5~commonly identifiable extrema per pulse (reference points P, Q, R, S, and T)  (see Fig.~\ref{pqrst}) though not all points can be easily identified on all pulses and the exact morphology can vary. We used freely available samples from the MIT-BIH Arrhythmia Database~\cite{PhysioNet}. \new{We only present our results over one sample (labeled ``100.dat'') since we found that results did not vary much between data samples.} These ECG recordings used a sampling rate of 360~Hz per channel with 11-bit resolution (see Fig.~\ref{basicecg}). We \new{keep the first} 4000~samples (11 seconds) and about 14 pulses, and we do no preprocessing such as baseline correction. We can estimate that a typical pulse has about 5 ``easily'' identifiable monotonic segments. Hence, out of 14 pulses, we can estimate that there are about 70 significant monotonic segments, some of which match the domain-specific markers (reference points P, Q, R, S, and T). A qualitative description of such data is useful for pattern matching applications.
 
The
running time as a function of $K$ is presented in Fig.~\ref{timevsk}. The scale-based segmentation implementation is faster than our implementations of the piecewise  linear
heuristics. On such a long
time series (4000~samples), our implementation of the bottom-up heuristic is much slower than the
alternatives.

{We want to
determine how well the  piecewise linear segmentation heuristics do comparatively. OMAFE is an absolute and not
relative error measure, but because the range
of the ECGs under consideration is roughly between 950 and 1150, we expect the OMAFE to never exceed 100 by much. }
The OMAFE with respect to the maximal number of segments ($K$) is given in Fig.~\ref{omafevsk}: it is a ``monotonicity spectrum.''
By counting on about 5~monotonic segments per pulse with a total of 14 pulses, there should about 70~monotonic segments in the 4000~samples under consideration.   We see that the decrease in OMAFE with the addition of new segments starts to level off between 50 and 70 segments as predicted. The addition of new segments past 70 ($K>70$) has little impact. The scale-based algorithm is optimal, but also at least 3 times more accurate than the top-down algorithm for larger $K$ and this is consistent over other data sets. In fact, the OMAFE becomes practically zero for $K>80$ whereas the OMAFE of the top-down linear regression algorithm remains at around 20, which is still significant.
The  bottom-up heuristic is more accurate than the top-down heuristic, but it still has about twice the OMAFE for large $K$. 
OMAFE of the scale-based algorithm is a non increasing function of $K$, a consequence of optimality.


\subsection{Temperature Recordings}

\begin{figure}
\subfigure[\label{figtemp1}Time Series]{
\centering\includegraphics[width=0.33\columnwidth]{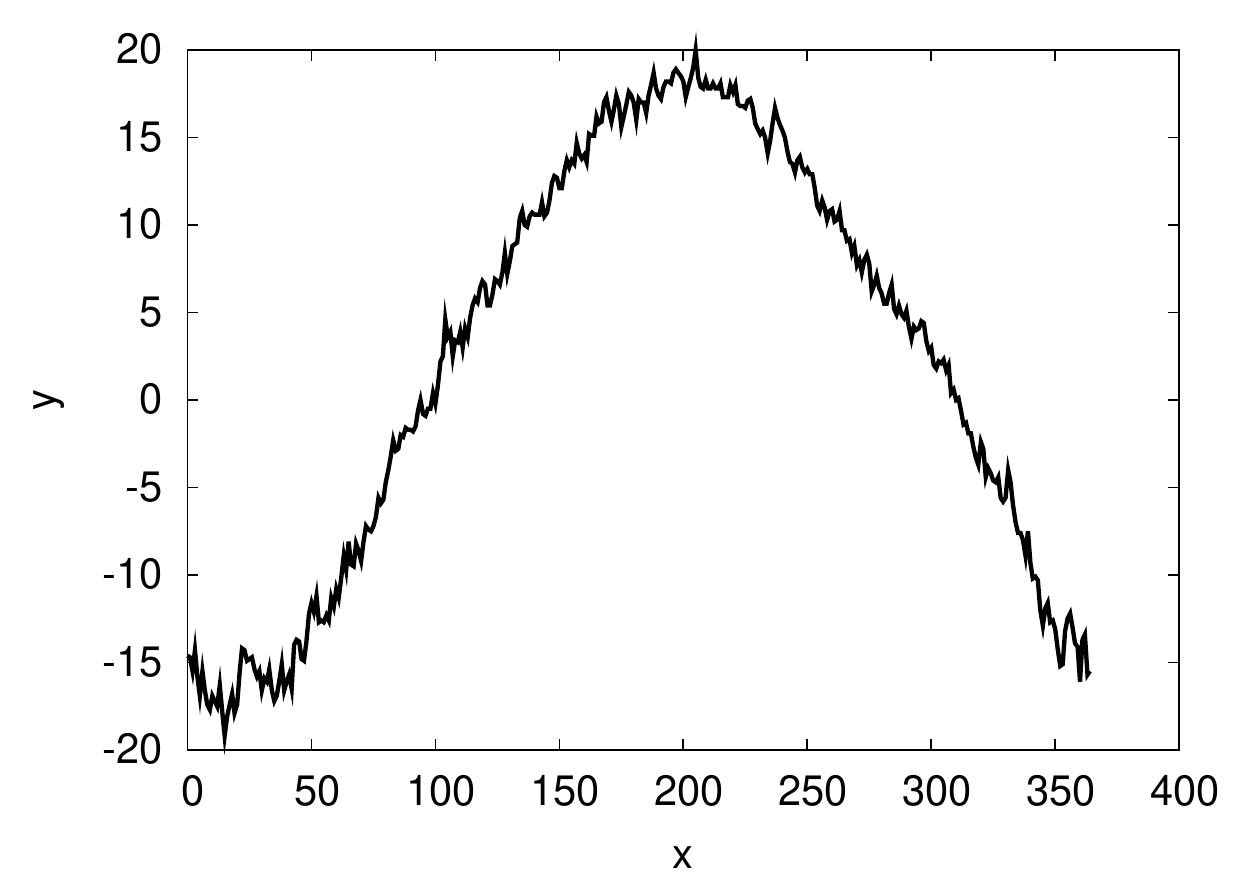}
}
\subfigure[\label{temp1timevsk}Time vs. number of segments $K$]{
\centering\includegraphics[width=0.33\columnwidth]{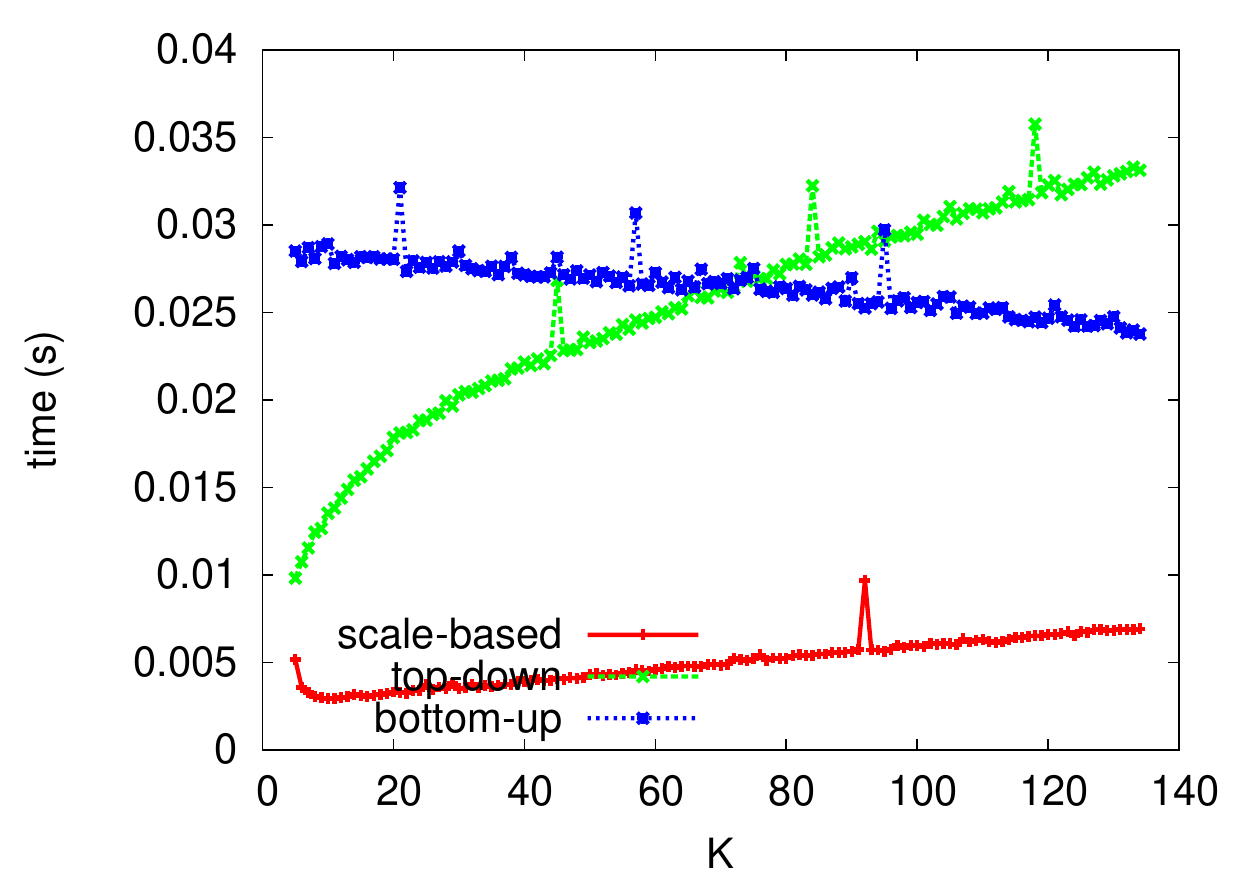}
}
\subfigure[\label{temp1omafevsk}OMAFE vs. number of segments $K$]{
\centering\includegraphics[width=0.33\columnwidth]{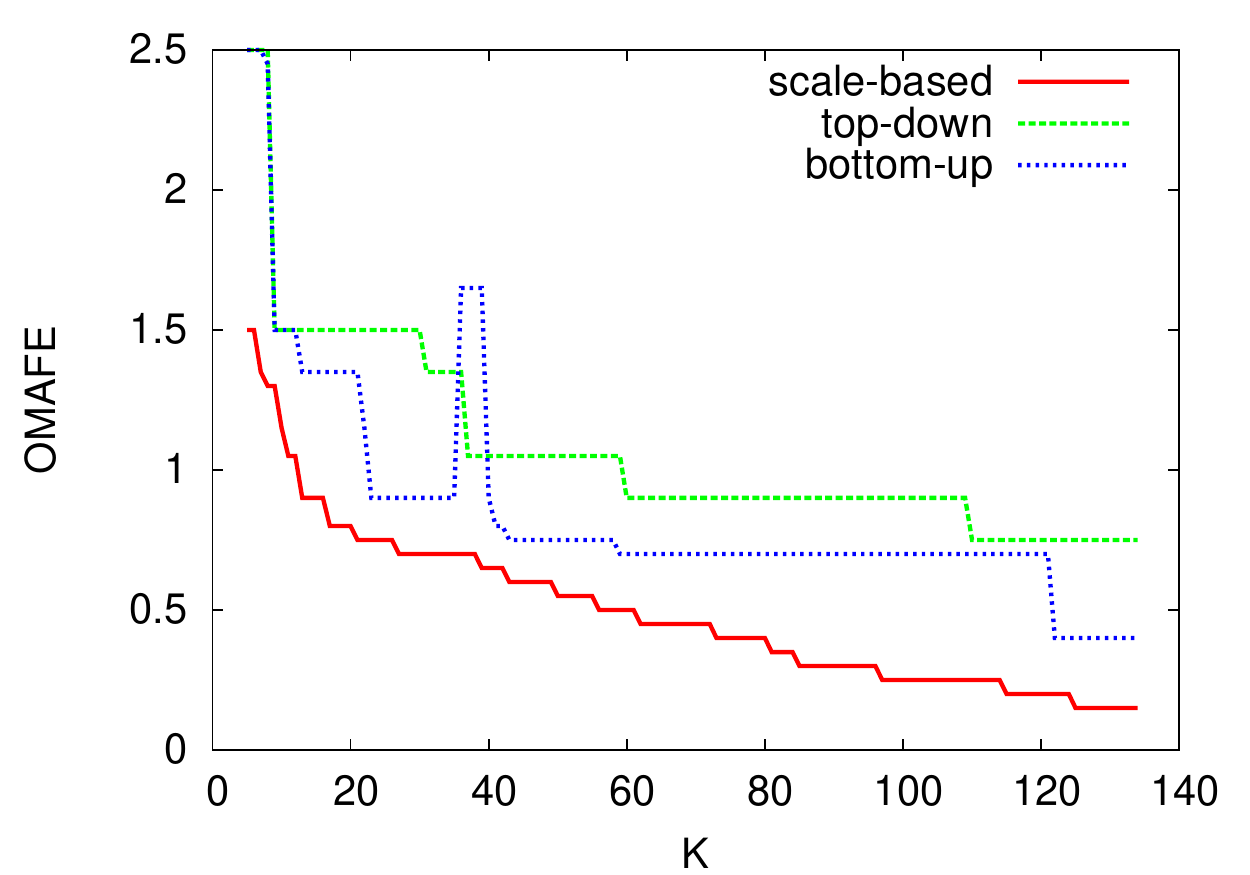}
}
\caption{Results of experiments over daily temperature data.}
\end{figure}

We consider the daily temperature recordings of the first of 35~weather stations in the MD*Base Daily temperature data set~\cite{mdbase}\footnote{the data is attributed to Ramsay and Silverman~\cite{ramsay1997afd}}. Since we only have one
year of recordings, only 365~data points are used (see Fig.~\ref{figtemp1}).
We also give the running times (see Fig.~\ref{temp1timevsk}) and the accuracy (see Fig.~\ref{temp1omafevsk}). Our implementation of the bottom-up heuristic is
now much faster due to small size of the times series, but the OMAFE, while superior to the top-down heuristic, exhibits a spurious spike near $K=40$,
showing the danger of relying on a piecewise linear heuristic to study the
monotonicity of a data set. Considering the OMAFE of our scale-based algorithm,
we notice that the accuracy increases slowly after $K=10$.

\subsection{Synthetic Random Walk Data}

\begin{figure}
\subfigure[\label{figbrownian}Time Series]{
\centering\includegraphics[width=0.33\columnwidth]{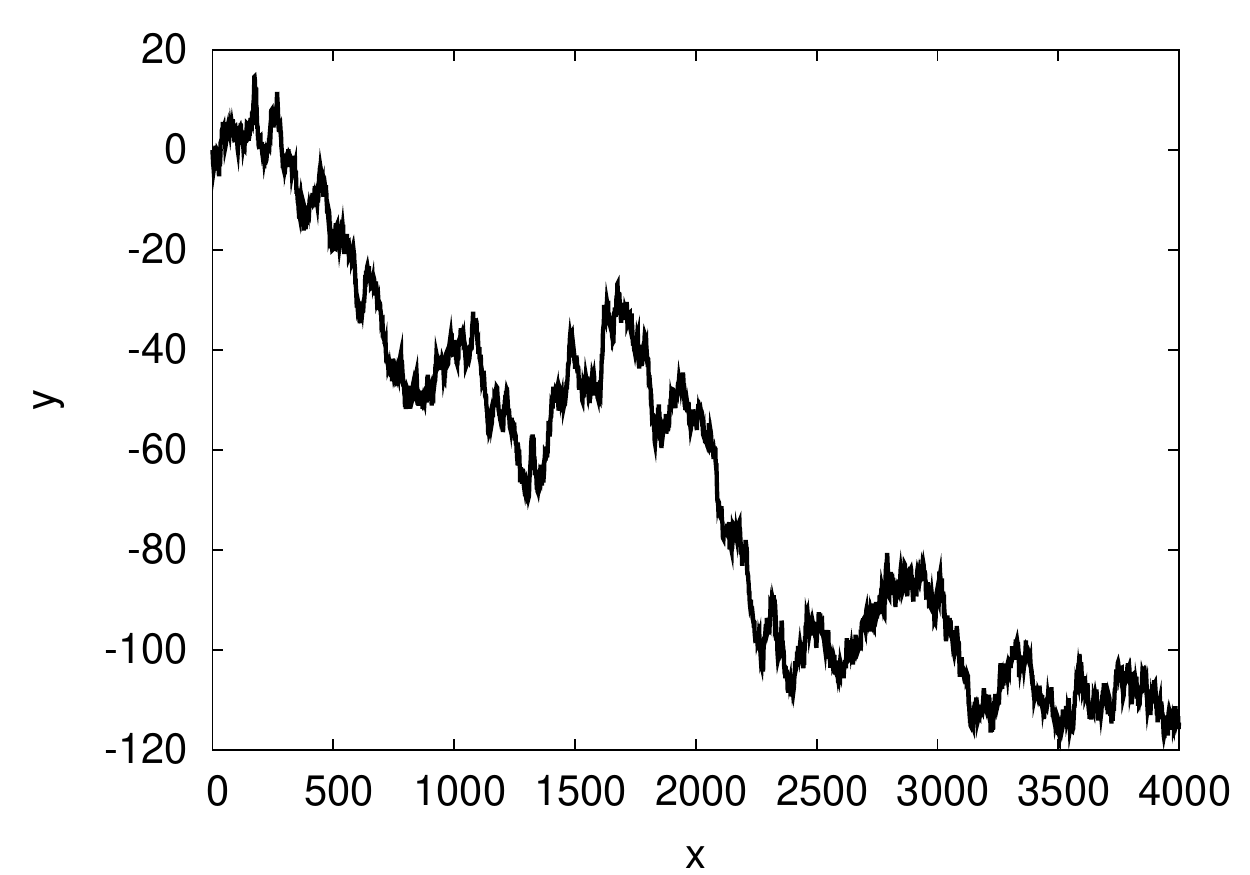}
}
\subfigure[\label{browniantimevsk}Time vs. number of segments $K$]{
\centering\includegraphics[width=0.33\columnwidth]{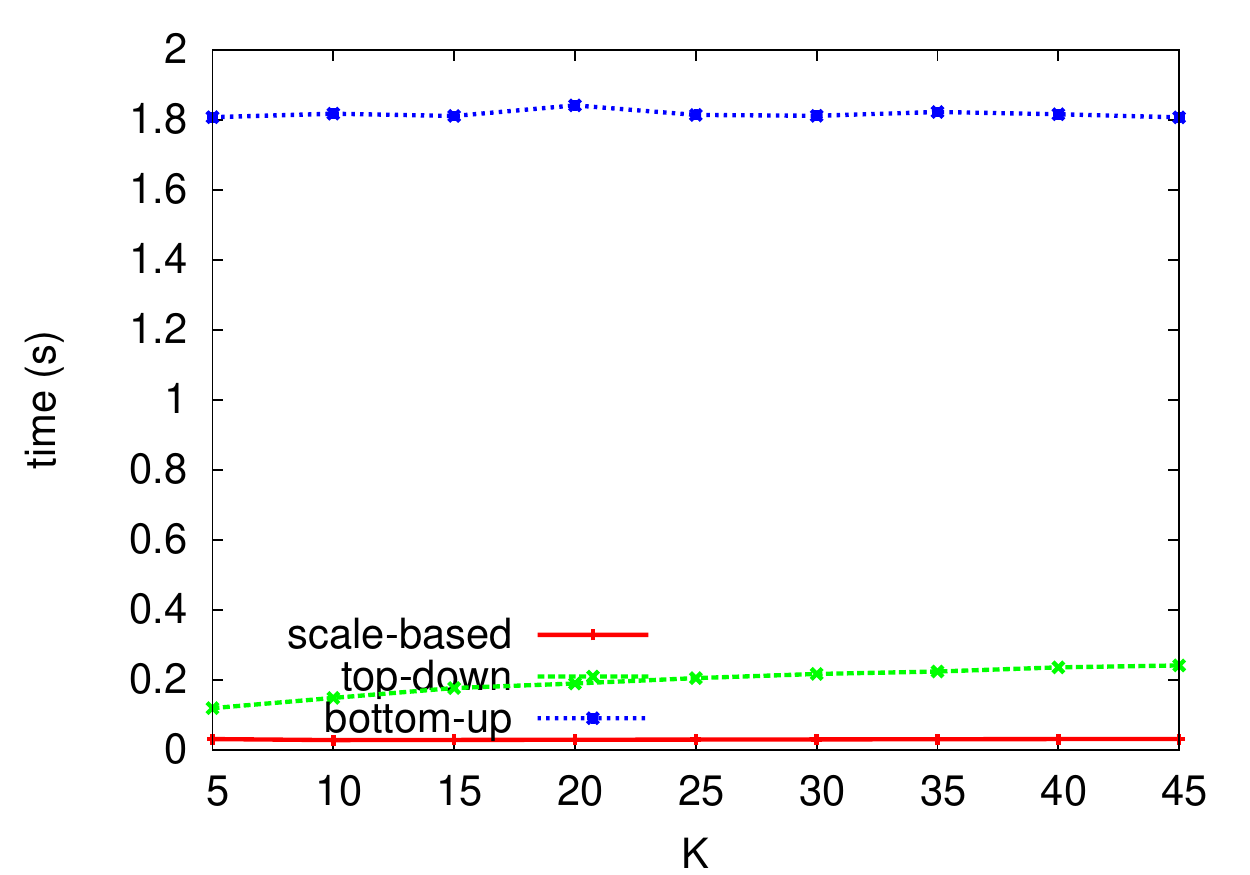}
}
\subfigure[\label{brownianomafevsk}OMAFE vs. number of segments $K$]{
\centering\includegraphics[width=0.33\columnwidth]{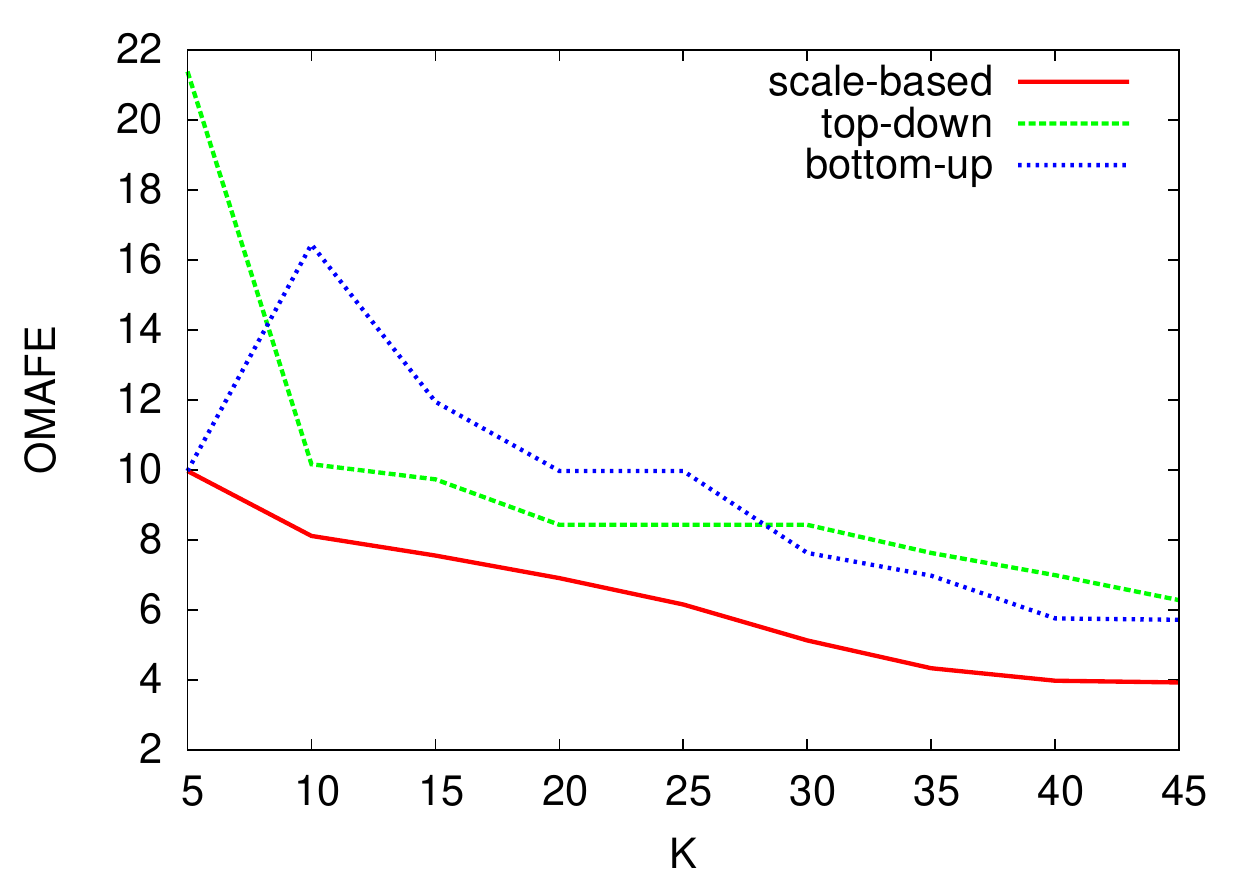}
}
\caption{Results of experiments over random walk.}
\end{figure}

Random walks are often used as models for
common time series such as stock prices. 
We generated a random walk $(i,y_i)_{i=1,\ldots,4000}$ using the formula $y_{i+1}=y_i+\epsilon$ where $\epsilon \sim N(0,1)$ (see Fig.~\ref{figbrownian}). The running times are nearly 
identical to the ECG case, as is expected since the time series have the
same length. However, the OMAFE differs (see Fig.~\ref{brownianomafevsk}):
using our optimal algorithm, the curve is smooth with no sharp drop. Meanwhile,
the bottom-up heuristic exhibits another spurious spike in the OMAFE (around $K=20$) while it provides the optimal segmentation at $K=5$.

\section{Conclusion and Future Work}

We presented optimal and fast algorithms to compute the best
piecewise monotonic segmentation in time $O(n)$ and the complete
OMAFE-versus-$K$ spectrum in time $O(n\log n)$.
Our experimental results suggest that one should be careful when deriving monotonicity information from piecewise linear segmentation heuristics.
Future work will focus on choosing the optimal number of segments for given applications.
We also plan to investigate the applications of the monotonicity spectrum as a robust
analysis. Further work to integrate flat segments is needed~\cite{YLBIJCAI05,LemireSDM2007}.

\bibliographystyle{ieeetr}
\bibliography{modelabstraction,yuhong,metric1d} 

\end{document}